\documentclass[12pt, a4paper]{article}
\usepackage[T1, T2A]{fontenc}
\usepackage[utf8]{inputenc}
\usepackage[russian, english]{babel}
\usepackage{amsmath, amssymb, amsthm}
\usepackage{geometry}
\usepackage{cases}
\usepackage{hyperref}
\usepackage{tikz-cd}
\newcommand{\bq}{\begin{equation}}
\newcommand{\eq}{\end{equation}}

\newtheorem{theorem}{Theorem}

\newtheorem{prop}{Proposition}
\newtheorem{st}{Statement}

\theoremstyle{definition}
\numberwithin{equation}{section}
\begin{document}


\title{On a Supersymmetric Generalization of the Matytsin Formula}

\date{}
\author{E. Porubov}
\maketitle
\begin{abstract}
We derive the leading asymptotics of the supersymmetric analogue of the Itzykson–Zuber integral for the supergroup $U(n|m)$ in the limit $n,m\to\infty$ when $n/m$ is kept finite. This generalizes the Matytsin result for $U(N)$. We show that it is described by the classical action of two independent Hamiltonian systems corresponding to the bosonic and fermionic sectors.
\end{abstract}

\section{Introduction}

Integrals over compact Lie groups play an important role in representation theory, random matrix theory, and mathematical physics. A fundamental result in this area is the Harish-Chandra formula \cite{HC}, which expresses integrals of exponentials of linear functions over a group in terms of a sum over the Weyl group and data defined on a Cartan subalgebra. That is, if $G$ is connected, compact and semisimple, $W$ is its Weyl group, $dg$ denotes the Haar measure, and $\phi,\chi\in\mathfrak{h}\subset\mathfrak{g}= \mbox{Lie}(G)$ lie in the Cartan subalgebra, then
\begin{equation*}\label{HCi}
       \int_G e^{\langle Ad_g \phi,\chi \rangle} dg=\frac{|W|^{-1}}{\Delta_\mathfrak{g}(\phi)\Delta_\mathfrak{g}(\chi)}\sum_{w\in W}\epsilon(w) e^{\langle w(\phi),\chi\rangle}\,.
\end{equation*}
Here $\langle .,.\rangle$ on $\mathfrak{g}$ is the Killing form, and $ \Delta_\mathfrak{g}(h):=\prod_{\alpha\in \Phi^+} \langle \alpha,h\rangle\,$.

In the case of the unitary group $U(N)$, this formula, after reducing the integral to eigenvalues of Hermitian matrices, leads to the determinantal representation of the Itzykson--Zuber integral \cite{IZ}. In mathematics it also known as the Harish-Chandra formula.
\begin{equation}\label{IZ}
I(\phi,\chi)=\int_{U(N)} \mathcal{D}U e^{N\,\text{Tr}(\phi U \chi U^{\dagger})}=\frac{\det(e^{N\phi_i \chi_j})}{\Delta(\phi)\Delta(\chi)}\,,
\end{equation}
where $\mathcal{D}U$ is the Haar measure on the group $U(N)$, $\phi,\chi$ are Hermitian $N\times N$ matrices with distinct eigenvalues $\phi_i,\chi_i$, and $\Delta(\phi)=\prod_{i<j}^N (\phi_i - \phi_j)$ is the Vandermonde determinant.

The Itzykson--Zuber integral arises in various problems of random matrix theory, in particular in the study of matrix models and eigenvalue statistics, as well as in theoretical physics, for example in the analysis of Yang--Mills theory on a cylinder \cite{GM}. On the one hand, it admits an explicit representation in terms of the Vandermonde determinant, while on the other hand, it exhibits nontrivial asymptotic behavior in the large-dimension limit. The study of this limit is closely related to large deviation–type problems and naturally leads to quantities of order $N^2$, interpreted as the free energy of statistical models.

Significant progress in understanding the asymptotics of the Itzykson--Zuber integral was achieved by A.~Matytsin in \cite{mat93}, where he showed that when $
N\to\infty$ the leading contribution can be described in terms of a variational problem for the eigenvalue density. Within this approach, the asymptotics has an exponential form with an exponent of order $N^2$, and the corresponding functional is expressed in terms of the classical action of a certain effective dynamical system. This result was subsequently established rigorously by A.~Guionnet and O.~Zeitouni in \cite{GZ}.

One of the key elements of this approach is the Hamiltonian interpretation of the resulting variational problem. The corresponding functional satisfies a Hamilton--Jacobi type equation, while the dynamics of the eigenvalue density is governed by a nonlinear complex Hopf equation. On the other hand, the Itzykson--Zuber integral is closely related to the characters of irreducible representations of the group $U(N)$ and, in particular, to Schur polynomials. This connection allows one to use its asymptotics to study the limiting behaviour of the corresponding representations, including the asymptotics of Young diagrams and the associated probability measures.

A natural direction for further development is the extension of this approach to a broader class of spherical integrals, including those associated with $\beta$-ensembles, as well as to supersymmetric generalizations arising in the context of Lie supergroups. In the latter case, the presence of both commuting and anticommuting variables plays a crucial role, leading to modifications of both the analytic and geometric aspects of the problem. Moreover, although the asymptotics of the Itzykson--Zuber integral has been understood for quite some time, the analysis of subleading corrections remains an open problem.

In this work we extend the Matytsin approach to the case of Lie supergroup $U(n|m)$. We derive the asymptotic of the Itzykson--Zuber type integral for the supergroup $U(n|m)$ in the limit $n,m\to\infty$ with a fixed ratio $n/m$.These results are derived following Matytsin arguments.The analytic justification of the collective-field limit is beyond the scope of this note.

We show that in this case the corresponding variational problem is given again by a Hamilton--Jacobi equation. The effective Hamiltonian system now decomposes into two parts corresponding to the bosonic and fermionic sectors. The group $U(n|m)$ has a subgroup $U(n)\times U(m)$ (the body of the supermanifold $U(m|n)$). The first factor we call the bosonic sector, the second factor we call the fermionic sector. The interaction between them manifests itself at the level of canonical transformations and leads to additional boundary contributions to the action.

The paper is organized as follows. In Section 2, we briefly overview the derivation of the Matytsin formula and discuss several technical refinements. In Section 3, we consider the generalization to $U(n|m)$ and derive the corresponding asymptotics. In the conclusion, we discuss possible directions for further research.

\section{The Matytsin formula for \(U(N)\)}

We briefly reproduce the arguments presented in \cite{mat93}. The main result is the description of the leading contribution to the asymptotic expansion of the integral (\ref{IZ}) as $N\to\infty$. For fixed eigenvalue densities $\rho(x),\sigma(x)$ of the Hermitian matrices $\phi,\chi$, the leading contribution is given by
\begin{equation*}
    I(\phi,\chi) \sim \exp{(N^2F_0[\rho,\sigma]+o(N^2))}\,,
\end{equation*}
where $\rho$ and $\sigma$ are the eigenvalue limiting empirical spectral densities of the prescribed source matrices $\phi$ and $\chi$, respectively. That is, the empirical measures $N^{-1}\sum_i \delta_{\phi_i}$ and $N^{-1}\sum_i \delta_{\chi_i}$ converge to compactly supported probability measures with densities $\rho,\sigma$ in the limit $N\to\infty$.
\begin{theorem}[\cite{mat93}, \cite{GZ}]
The functional $F_0[\rho,\sigma]$ is given by
\begin{equation}\label{ans}
    F_0[\rho,\sigma]=S[\rho,\sigma]+\frac{1}{2}\int dx(\rho(x)+\sigma(x))x^2-\frac{1}{2}\int dxdy\,(\rho(x)\rho(y)+\sigma(x)\sigma(y))\ln{|x-y|}\,.
\end{equation}
The corresponding functional $S[\rho,\sigma]$ is the classical action evaluated along a trajectory determined by the Hamiltonian
\begin{equation*}
    H[q,p]=\frac{1}{2}\int dx\, q(x)\left\{\left(\frac{\partial p(x)}{\partial x} \right)^2-\frac{\pi^2}{3}q^2(x)\right\}\,,
\end{equation*}
where $q,p$ are the canonical coordinate and momentum satisfying Poisson bracket $\{q(x),p(y)\}=\delta(x-y)$, with the boundary conditions
\begin{equation*}
    q(x,t=0)=\rho(x)\,,\quad q(x,t=1)=\sigma(x)\,.
\end{equation*}
\end{theorem}

The proof proceeds as follows: consider a one-parameter family of integrals
\begin{equation*}
    I(\phi,\chi | t)=\frac{\Delta^2(\chi)}{(2\pi t)^{\frac{N^2}{2}}}\int \mathcal{D}U e^{-\frac{N}{2t}\text{Tr}(\phi-U\chi U^{\dagger})^2}=\frac{\Delta^2(\chi)}{(2\pi t)^{\frac{N^2}{2}}}e^{-\frac{N}{2t}\text{Tr}(\phi^2 + \chi^2)}I\left(\frac{\phi}{\sqrt{t}},\frac{\chi}{\sqrt{t}}\right)\,,
\end{equation*}
which satisfies a heat-type equation
\begin{equation}\label{heat}
    2N\frac{\partial I(\phi,\chi | t)}{\partial t}=\frac{1}{\Delta(\phi)}\sum_{i=1}^N \frac{\partial^2}{\partial \phi_i^2}\Delta(\phi)I(\phi,\chi | t)\,.
\end{equation}
And $I(\phi,\chi|t\to0)=\sum_{\sigma\in S_N}\delta(\phi-\chi_\sigma)$. To obtain the explicit form of $F_0[\rho,\sigma]$, one needs to consider the limit of the solution to this equation with the ansatz $I(\phi,\chi|t)=\exp{(N^2 W[\rho,\sigma|t])}$. Below we suppress the dependence of the functional $W$ on the eigenvalue densities and the parameter $t$. Then equation (\ref{heat}) can be rewritten as
\begin{equation*}
    2\frac{\partial}{\partial t}W=\frac{1}{N}\sum_{i=1}^N\frac{\partial^2}{\partial \phi_i^2} W+\frac{1}{N^3}\sum_{i=1}^N\frac{\partial^2}{\partial \phi_i^2}\ln{\Delta(\phi)}+N\sum_{i=1}^N\left[\frac{\partial}{\partial \phi_i}\Big(W+\frac{\ln{\Delta(\phi)}}{N^2}\Big) \right]^2\,,
\end{equation*}
where $\ln \Delta(\phi)=\frac{1}{2}\sum_{i\neq j}^N \ln|\phi_i-\phi_j|$. Since $\rho(x)=\frac{1}{N} \sum_{i=1}^N \delta(x-\phi_i)$, straightforward computations yield
\begin{equation*}
    \frac{\partial W}{\partial \phi_i}=\frac{1}{N}\left(\frac{\partial}{\partial x}\frac{\delta W}{\delta \rho(x)} \right)\Bigg|_{x=\phi_i}\,.
\end{equation*}
The first term is therefore of order $\mathcal{O}(1/N)$ and can be neglected, whereas the second term contributes at order $\mathcal{O}(1)$ and cannot be discarded.
The $\mathcal{O}(1)$ contribution in the second term can be immediately seen from the harmonicity of the Vandermonde determinant
\begin{equation*}
    \sum_{i=1}^N\frac{\partial^2}{\partial \phi_i^2}\Delta(\phi)=0\implies \sum_{i=1}^N\frac{\partial^2}{\partial \phi_i^2}\ln \Delta(\phi)+\sum_{i=1}^N\left[\frac{\partial}{\partial \phi_i}\ln \Delta(\phi)\right]^2=0\,,
\end{equation*}
where in the original paper the expression $\frac{\partial}{\partial \phi_i}\ln \Delta(\phi)=V(\phi_i)$ plays a central role in describing the asymptotics. Following the same reasoning as in Matytsin’s work, the $\mathcal{O}(1)$ contribution in the second term, in the limit $N\to\infty$, is given by
\begin{equation*}
   \frac{-1}{N^3}\sum_{i=1}^N\left[\frac{\partial}{\partial \phi_i}\ln \Delta(\phi)\right]^2=-\frac{1}{N^3}\sum_{i\neq j}^N\frac{1}{(\phi_i-\phi_j)^2}\cong-\frac{1}{N}\sum_{i\neq j}^N \frac{\rho^2(\phi_i)}{(i-j)^2}\overset{N\to\infty}{\longrightarrow} -\int dx\, \rho(x)^3\frac{\pi^2}{3}\,.
\end{equation*}
Here we employ a non-rigorous justification of the limit, assuming that our analysis is performed in a neighbourhood of the saddle point of the original integral. In this case, the leading contribution is obtained by using the approximation $|\phi_i-\phi_j|\cong|i-j|/(N\rho(\phi_i))$. Thus, the $\mathcal{O}(1)$ part of equation (\ref{heat}) in the limit $N\to\infty$ takes the form
\begin{equation*}
    2\frac{\partial}{\partial t}W=-\int dx\, \rho^3(x)\frac{\pi^2}{3}+\int dx\,\rho(x) \left\{\frac{\partial}{\partial x}\left[\frac{\delta W}{\delta \rho(x)}+\int dy \rho(y)\ln |x-y| \right] \right\}^2\,.
\end{equation*}
\begin{prop}
This equation can be represented as a Hamilton--Jacobi equation
\begin{equation*}
    \frac{\partial W}{\partial t}+H\left[\rho(x),\frac{\delta W}{\delta \rho(x)}\right]=0\,,
\end{equation*}
where $t\to-t$ and the corresponding Hamiltonian is given by
\begin{equation}\label{dh}
    H[q,p]=\frac{1}{2}\int dx\, q(x)\left\{\left(\frac{\partial}{\partial x}\left[p(x)+\int dy q(y)\ln|x-y| \right]\right)^2-\frac{\pi^2}{3}q^2(x) \right\}\,,
\end{equation}
where $p$ is the canonical momentum conjugate to $q$ 
\[
\{q(x),p(y)\}=\delta(x-y)\,,\quad\{q(x),q(y)\}=\{p(x),p(y)\}=0\,.
\]
\end{prop}
Note that the change of variables $q(x)\to q(x)$, $p(x)\to p(x)+\int dy\, q(y)\ln |x-y|$ is a canonical transformation, i.e., a map preserving the Poisson brackets. Therefore, the Hamiltonian
\begin{equation}\label{ham}
    H[q,p]=\frac{1}{2}\int dx \,q(x)\left\{\left(\frac{\partial p(x)}{\partial x}\right)^2 -\frac{\pi^2}{3}q^2(x)\right\}
\end{equation}
generates the same Hamiltonian system as (\ref{dh}). This Hamiltonian was obtained in the work of A.~Matytsin and in \cite{GZ} the momentum with the logarithmic correction was written down. The equations of motion are related to the complex Hopf equation after the substitution $f(x,t)=\partial_x p(x,t)+i\pi q(x,t)$.
\begin{align*}
    &\frac{\partial}{\partial t} f(x,t)+f(x,t) \frac{\partial}{\partial x} f(x,t)=0\,,\\
    &\left\{
    \begin{array}{cc}
         \mbox{Im }f(x,t=0)=\pi\rho(x)\,,&  \\
         \mbox{Im }f(x,t=1)=\pi \sigma(x)\,.& 
    \end{array}
    \right.
\end{align*}
Since a canonical transformation adds a boundary contribution to the action being evaluated, one has
\begin{equation*}
    W[\rho,\sigma]=S_0[\rho,\sigma]+G[q]\Big|^{t=1}_{t=0}\,,\quad G[q]=\frac{1}{2}\int dx dy\,q(x,t)q(y,t)\ln|x-y|\,,
\end{equation*}
where the functional $S_0[\rho,\sigma]$ is the classical action for the system with Hamiltonian (\ref{ham}), and $G[q]$ is the generating functional of the canonical transformation. 

It remains to verify that this action describes the asymptotics of the Itzykson--Zuber integral (\ref{IZ}). To this end, following \cite{mat93}, we deduce from the identity
\[
D_kI(\phi,\chi)=\operatorname{Tr}(\chi^k)I(\phi,\chi)\,,\quad D_k=\frac{1}{\Delta(\phi)}\frac{1}{N^k}\sum_{i=1}^N\frac{\partial^k}{\partial \phi_i^k}\Delta(\phi)
\]
that the function
\[
H_k(p,q,t)=\frac{1}{2\pi i t(k+1)}\int dx\,\left\{\left(x+(\partial_xp+i\pi q)t\right)^{k+1}-\left(x+(\partial_xp-i\pi q)t\right)^{k+1}\right\},\, k\in \mathbb{N}\,,
\]
must satisfy
\[
H_k(p,q,t=1)=\int x^k\sigma(x) dx\,.
\]
This follows by applying Matytsin’s conservation law to the time-reversed trajectory $\dot H_k=0$.

Thus, the final expression for the leading contribution to the asymptotics of the integral (\ref{IZ}) is precisely (\ref{ans}), which completes the derivation of the asymptotic formula.

Furthermore, the determinantal representation of the Itzykson--Zuber integral is closely related to the Weyl character formula. Recall that the character of an irreducible representation $\pi$ of a group $G$ with highest weight $\lambda$ is defined as
\[
\text{ch}_\pi(h)=\frac{\sum_{\sigma\in W}\varepsilon(\sigma)e^{\sigma(\lambda+\rho)(h)}}{\sum_{\sigma\in W}\varepsilon(\sigma)e^{\sigma(\rho)(h)}}=\frac{\sum_{\sigma\in W}\varepsilon(\sigma)e^{\sigma(\lambda+\rho)(h)}}{\prod_{\alpha\in \Phi^+}(e^{\frac{\alpha(h)}{2}}-e^{-\frac{\alpha(h)}{2}})}\,,
\]
here $W$ is the Weyl group, $h\in \mathfrak{h}\subset \text{Lie}(G)$ lies in the Cartan subalgebra, and $\rho$ is the half-sum of positive roots. In the case of the group $U(N)$, fixing an element in the Cartan subalgebra $h=(i\phi_1,i\phi_2,\dots,i\phi_N)$, this formula takes the form
\[
s_\lambda(e^{i\phi})=\text{ch}_\lambda(e^{i\phi_1},e^{i\phi_2},\dots,e^{i\phi_N})=\frac{\det\left(e^{i\phi_i x_j}\right)}{\prod_{i<j}\left(e^{i\phi_i}-e^{i\phi_j}\right)}\,,\quad x_j=N\left(\frac{\lambda_j}{N}+1-\frac{j}{N}\right)\,,
\]
where $\cdots \leq \lambda_{i+1}\leq\lambda_i\leq\lambda_{i-1}\leq\cdots$ are the components of the highest weight $\lambda$. Denote by $\chi_j=\lambda_j/N+1-j/N$ the shifted weights and assume that in the large-$N$ limit they are described by a smooth function $\sigma$, i.e., $N^{-1}\sum_i \delta_{\chi_i}$ converges to a compactly supported probability measure $\mu$ with density $\sigma$. The relation between the asymptotics of characters of irreducible representations of $U(N)$ and the Itzykson--Zuber integral is discussed in \cite{GM}, \cite{Zel} and requires a separate analysis. We only note that in the finite-dimensional case the character $\text{ch}_\lambda(i\phi)$ can be obtained by analytic continuation of one of the parameters in the Itzykson--Zuber integral
\[
\det(e^{Na_ib_j})\quad\overset{a_j\,\to\, ia_j}{\longrightarrow} \quad\Delta(e^{ib})\text{ch}_{a}(b)\,,
\]
under the identification $a_i=\chi_i$, $b_i=\phi_i$.

On the other hand, for the specialization of the Schur polynomial $s_\lambda(e^\phi)$ with $\phi_i\in\mathbb{R}$, one immediately sees that
\[
s_\lambda(e^\phi)=I(\phi,\chi)\Delta(\chi)\frac{\Delta(\phi)}{\Delta(e^\phi)}\,,
\]
where $\phi=\text{diag}(\phi_1,\dots,\phi_N)$ and $\chi$ is related to the partition $\lambda$ as above. Thus, the asymptotics of the Itzykson--Zuber integral provides a way to describe the asymptotics of Schur polynomials under real specialization.

\section{The Itzykson-Zuber type integral for \(U(n|m)\) and its asymptotics}

A natural direction for further development is the extension of this approach to the supersymmetric setting, specifically to the analogues of the Itzykson-Zuber integral over the Lie supergroup $U(n|m)$. To do this rigorously, let us introduce the underlying supergeometry. We first define the \(U(n|m)\) as a real form of $GL(n|m)$ and the normalization of the invariant Berezin integral. We then use the radial differential equation to derive the collective-field description under the continuum assumptions stated below. For fundamental concepts regarding supermanifolds and supergroups, we refer to the classic works of Berezin \cite{ber}, Leites \cite{Leit}, Deligne and Morgan \cite{DM} and Varadarajan \cite{Var}. An accessible exposition for physicists can be found in \cite{efetov}.

\subsection{\(GL(n|m)\)}
A complex vector space $V$ is a $\mathbb{Z}_2$-graded vector space of dimension $p|q$ if it decomposes as $V=V_0\oplus V_1$, with $\dim V_0=p$, $\dim V_1 =q$. We introduce the parity operator $\alpha$, such that $\alpha(v)=i$ for $v\in V_i$. The homomorphisms of $\mathbb{Z}_2$-graded vector spaces $\text{Hom}_{\text{Vect}} (V,W)$ are linear maps, and we also have $\text{Hom}_{\text{Vect}}(V,V)=\text{End}_{\text{Vect}}(V)$. Clearly, $\mathbb{Z}_2$-graded vector spaces induce a $\mathbb{Z}_2$-grading on $\operatorname{Hom}$.
Consider the $\mathbb{Z}_2$-graded vector space $\mathbb{C}^{p|q}$ and fix homogeneous basis $\{e_i\}$ in which the first $p$ vectors form a basis of the even part, while the next $q$ vectors form a basis of the odd part. Such a basis will be called standard. Linear operators in the standard basis $A\in\text{End}_{\text{Vect}}(\mathbb{C}^{p|q})$ are represented by block matrices
\begin{equation*}
    A = \begin{pmatrix} A_{11} & A_{12} \\ A_{21} & A_{22} \end{pmatrix}=\begin{pmatrix}
        A_{11}& 0\\ 0 & A_{22}
    \end{pmatrix}+\begin{pmatrix} 0 & A_{12} \\ A_{21} & 0 \end{pmatrix}:  \mathbb{C}^{p|q} \longrightarrow \mathbb{C}^{p|q}\,.
\end{equation*}
Where the first term is the even part of the operator, and the second is the odd part.

To upgrade this $\mathbb{Z}_2$-graded space to a super vector space, one must equip its tensor category with the Koszul sign rule
\begin{align*}
    c_{V,W}:\, V\otimes W&\to W\otimes V\,,\\
    v\otimes w\,\,&\mapsto (-1)^{\alpha(v)\alpha(w)}w\otimes v\,,
\end{align*}
for homogeneous vectors $v,w$. This tensor category is called the category of supervector spaces, with objects being supervector spaces. The morphisms in such a category are linear maps that preserve the $\mathbb{Z}_2$ grading. Now we will consider $\mathbb{C}^{p|q}$ as a supervector space. Further on, we will write $\text{Hom}$, meaning $\text{Hom}_{\text{sVect}}$. Note that if we add operators that change parity to $\operatorname{Hom}(V,W)$ as homomorphism of supervector spaces, we will obtain a supervector space, which we will denote by $\underline{\operatorname{Hom}}(V,W)$.


Let $\Lambda$ be a complex Grassmann algebra with generators $\{\xi_i\}_{i=1}^{2l}$, that is, $\Lambda$ is the quotient algebra of the free associative algebra over $\mathbb{C}$ by the relations $\xi_i\xi_j+\xi_j\xi_i=0$. Define $\alpha(\xi_i)=1$, which induces a grading $\Lambda=\Lambda_0\oplus\Lambda_1$ and endows $\Lambda$ with the structure of a $\mathbb{Z}_2$-graded space. Multiplication in the $\mathbb{Z}_2$-graded space $\Lambda$ is consistent with the sign rule, which makes $\Lambda$ a commutative superalgebra.

Let us consider the free $\Lambda$-module $\Lambda^{p|q}$, that is, a free module as an ungraded module with a homogeneous basis, i.e., $\Lambda^{p|q}=\mathbb{C}^{p|q}\otimes_\mathbb{C}\Lambda$, where $\mathbb{C}^{p|q}$ and $\Lambda$ are understood as super vector spaces.
We denote the even part by $\mathbb{C}^{p|q}(\Lambda):=\Lambda^{p|q}_0$ which is referred to in \cite{ber} as the Grassmann envelope.


The space $\operatorname{End}(\Lambda^{p|q})=(\underline{\operatorname{End}}(\mathbb{C}^{p|q})\otimes_\mathbb{C} \Lambda)_0$ is the Grassmann envelope of  $\underline{\operatorname{End}}(\mathbb{C}^{p|q})$, which we will denote as $\operatorname{End}(p|q)(\Lambda)$.
The elements of $\text{End}(p|q)(\Lambda)$ are the matrices in homogeneous basis of the form
\begin{equation*}
    A = \begin{pmatrix} A_{11} & A_{12} \\ A_{21} & A_{22} \end{pmatrix}\,,\qquad \text{matrix elements }(A_{11})_{ij},(A_{22})_{ij}\in \Lambda_0\,,\quad (A_{12})_{ij},(A_{21})_{ij}\in \Lambda_1\,.
\end{equation*}

The superanalogue of the trace  $\operatorname{STr}:\operatorname{End}(p|q)(\Lambda)\to \Lambda_0$ with the defining relation $\operatorname{STr}(AB)=\operatorname{STr}(BA)$ for $A,B\in \operatorname{End}(p|q)(\Lambda)$ along with $\Lambda_0$ linearity in homogeneous basis is
\begin{equation*}
    \operatorname{STr}(A)=\operatorname{Tr}(A_{11})-\operatorname{Tr}(A_{22})\,.
\end{equation*}
Define the map $m:\Lambda\to \mathbb{C}$ which acts as follows $m(\sum_I f_I \xi^I)=f_0\in\mathbb{C}$.
Let us write an element $A=A_0+N\in \operatorname{End}(\mathbb{C}^{p|q})(\Lambda)$, where $A_0=m(A)$ is the numerical part and $N$ is nilpotent part. Then the element $A$ is invertible if and only if $A_0$ is invertible. Denote by $\text{GL}(p|q)(\Lambda)$ the group of invertible elements in the Grassmann envelope $\text{End}(p|q)(\Lambda)$. Define the homomorphism
\[
\operatorname{Ber}: \text{GL}(p|q)(\Lambda)\longrightarrow \text{GL}(1|0)(\Lambda)=\Lambda_0^\times\,,
\]
That means $\operatorname{Ber}(AB)=\operatorname{Ber}(A)\operatorname{Ber}(B)$. As a superanalog of the determinant, characterized, analogously to determinant by its linearization: for $A\in \operatorname{End}(p|q)(\Lambda)$ and even formal parameter $\varepsilon$ such that $\varepsilon^2=0$
\[
\operatorname{Ber}(1+\varepsilon A)=1+\varepsilon\operatorname{STr}(A)\,.
\]
In particular, if $A=\begin{pmatrix}
    A' & 0\\ 0 & A''
\end{pmatrix}$, then $\operatorname{Ber}(1+\varepsilon A)=\operatorname{det}(1+\varepsilon A')\operatorname{det}(1+\varepsilon A'')^{-1}$. If $\exp(A)$ converges for $A\in \operatorname{End}(p|q)(\Lambda)$, then
\begin{equation*}
    \operatorname{Ber}(e^A) = e^{\operatorname{STr}(A)}\,.
\end{equation*}
The Berezinian in the standard basis is given by
\begin{equation*}
    \operatorname{Ber}(A) = \det(A_{11} - A_{12}A_{22}^{-1}A_{21})\det(A_{22}^{-1})\,.
\end{equation*}

\subsection{\(U(n|m)\)}
Consider the Grassmann algebra $\Lambda$ with an even number of generators and an antilinear anti-automorphism $*:\Lambda\to\Lambda$ such that, for any homogeneous elements $f,g\in \Lambda$ 
\begin{align*}
    i)&\quad (fg)^*=(-1)^{\alpha(f)\alpha(g)}g^*f^*\,,\\
    ii)&\quad (f^*)^*=(-1)^{\alpha(f)}f\,.
\end{align*}
For such an anti-automorphism, there always exists generators $\{\xi_1,...,\xi_l,\xi_1^*,...,\xi_l^*\}$, such that 
\[
(\xi_i)^*=\xi_i^*\,,\quad (\xi_i^*)^*=-\xi_i\,.
\]
See, e.g., \cite{ber}. We introduce a Hermitian inner product on the space $\mathbb{C}^{p|q}(\Lambda)$ with values in $\Lambda_0$
\begin{equation*}
    (x,y)=\sum_{k}x^*_ky_k\,,\quad\text{where}\quad x=\sum_kx_ke_k\,,y=\sum_k y_ke_k\,.
\end{equation*}
Here $\{e_i\}$ is a homogeneous basis of the super vector space $\mathbb{C}^{p|q}$.
The subgroup of $\operatorname{GL}(p|q)(\Lambda)$ preserving this inner product is denoted by $U(p|q)(\Lambda)$.

Using the anti‑automorphism $*$ fixed on $\Lambda$, the operations $*$ and $\dagger$, analogous to complex conjugation and Hermitian conjugation, respectively, can be defined on $\operatorname{End}(p|q)(\Lambda)$
\begin{equation*}
A=(A_{ik})\,\quad A^*=(A^*_{ik})\,,\quad A^\dagger=(A^*_{ki}(-1)^{\alpha(i)(\alpha(k)+1)})\,, \quad \text{where } \alpha(i) \text{ is a parity of }e_i
\end{equation*}
and then $A\in U(p|q)(\Lambda)$ if and only if $A^\dagger A=1$.

In order to define complex Lie supergroups rigorously, consider an open subset $U\subset \mathbb{C}^p$ and the sheaf of supercommutative rings $\mathcal{O}_U^{p|q}$ defined by
\begin{equation*}
    \mathcal{O}_U^{p|q}:V\longrightarrow \mathcal{O}_{\text{hol}}(V)\otimes_\mathbb{C}\Lambda\,,\qquad V\subset U\quad\text{open}\,,
\end{equation*}
where $\mathcal{O}_{\text{hol}}(V)$ is the algebra of holomorphic functions on $V$ and $\Lambda$ is the Grassmann algebra with $q$ generators. A complex analytic supermanifold of dimension $p|q$ is a locally ringed superspace $(M, \mathcal{O}_M)$ that is locally isomorphic to $\left(U,\mathcal{O}_U^{p|q} \right)$. Morphisms between such supermanifolds are morphisms of the corresponding locally ringed superspaces. A complex analytic Lie supergroup is a complex analytic supermanifold equipped with a multiplication morphism $m:G\times G\to G$, together with a unit $e:\mathbb{C}^{0|0}\to G$ and an inversion $i:G\to G$ satisfying the standard commutative diagrams. A natural example is the Lie supergroup $\operatorname{GL}(p|q)$, regarded as an open subsupermanifold of $\mathbb{C}^{p^2+q^2|2pq}$ and, as a topological space, isomorphic to $\operatorname{GL}(p)\times \operatorname{GL}(q)$. With respect to the Hermitian structure introduced above, the unitary Lie supergroup $U(p|q)$ is a real form of $\operatorname{GL}(p|q)$.

Note that in the reasoning above, the Grassmann algebra can be replaced with another supercommutative superalgebra $\mathcal{A}$ and we can define $\operatorname{GL}(p|q)(\mathcal{A})$ as $\mathcal{A}$ point of the supergroup $\operatorname{GL}(p|q)$ in the sense of the point functor.



\subsection{The asymptotics of the Itzykson-Zuber type integral}

Thus, one can consider a supersymmetric analogue of the Itzykson--Zuber integral over the Lie supergroup $U(n|m)$. 
Let $\phi,\chi\in i\mathfrak{h}_0$, where $\mathfrak{h}_0$ is the Cartan subalgebra $\mathfrak{h}_0\subset \mathfrak{u}(n)\oplus\mathfrak{u}(m)\subset\operatorname{Lie}(U(n|m))$. This means that $\phi$ and $\chi$ are diagonal real matrices with simple spectrum. Here we do not consider the case of a supermatrix model, where $\phi,\chi$ are super Hermitian matrices.
\begin{prop}\label{detf}
Then one has
\begin{equation*}
I_s(\phi,\chi)=\underset{U(n|m)}{\int} \mathcal{D}Ue^{n\text{STr}(\phi U \chi U^\dagger)}=\frac{\det(e^{n\phi_i\chi_j})_{i,j=1}^n\det(e^{-n\bar \phi_i\bar\chi_j})_{i,j=1}^m}{\Delta_s(\phi)\Delta_s(\chi)},
\end{equation*}
\begin{equation*}
    \text{where}\quad \Delta_s(\phi):=\frac{\prod_{i<j}^n (\phi_i-\phi_j)\prod_{i<j}^m (\bar\phi_i-\bar\phi_j)}{\prod_{i=1}^n\prod_{j=1}^m(\phi_i-\bar\phi_j)},\quad \phi_{i+n}=\bar\phi_i, \chi_{i+n}=\bar \chi_i\,,
\end{equation*}
and we assume that all combinatorial constants appearing in front of the determinants are absorbed into the measure $\mathcal{D}U$. This measure can be understood as two Haar measures on $U(m) \times U(n)$ and a Berezin integral over $2mn$ variables (see \cite{invint} and appendix).
\end{prop}
This result, in the context of supermatrix models, was obtained in \cite{Umn}. For the reader's convenience, we outline the proof in the appendix. 

As in the classical case, we introduce a one-parameter family of integrals a one-parameter family of integrals
\begin{align}\label{sheat}
  I_s(\phi,\chi|t):=&\frac{\Delta_s^2(\chi)}{ t^{\frac{(n-m)^2}{2}}}\int DUe^{-\frac{n}{2t}\text{STr}(\phi-U\chi U^\dagger)^2}=\frac{\Delta_s^2(\chi)}{ t^{\frac{(n-m)^2}{2}}}e^{-\frac{n}{2t}\text{STr}(\phi^2+\chi^2)}I_s(\frac{\phi}{\sqrt{t}},\frac{\chi}{\sqrt{t}})\,.\notag
\end{align}
It also satisfies a heat-type equation
\begin{equation}\label{she}
    2n\frac{\partial}{\partial t}I_s(\phi,\chi|t)=\frac{1}{\Delta_s(\phi)}\sum_{i=1}^{m+n}(-1)^{p(\phi_i)}\frac{\partial^2}{\partial \phi_i^2}\Delta_s(\phi)I_s(\phi,\chi|t)\,,
\end{equation}
where $p(\phi_i)$ denotes the label of the element $\phi_i$, in the sense that it equals $0$ for $\phi_i$ and $1$ for $\bar{\phi}_i$. Since, as in the classical case, the eigenvalues lie on the real line, we may similarly introduce eigenvalue densities for the boson–boson block $\phi_i$ and the fermion–fermion block $\bar{\phi}_i$ as $\rho$ and $\bar{\rho}$, respectively. As in the classical case, we assume that the spectra are non-degenerate and that the densities are smooth functions with compact support. As in Matytsin’s approach, we assume a regular continuum limit in which the discrete logarithmic sums admit the density representations used below. This assumption also applies to the mixed sums involving the two spectral blocks. Overlapping limiting supports are allowed. Similarly, $\sigma,\bar \sigma$ are the density of $\chi$ and $\bar\chi$, respectively. The asymptotics in this approach is obtained analogously to the classical case if we consider the limit $n,m\to\infty$ with $n/m=\text{const}$ and use the ansatz $I_s(\phi,\chi|t)=\exp{n^{2}W_s[\rho,\bar\rho,\sigma,\bar\sigma|t]}$. In other words, we assume that the limit exists and is smooth. In this setting, the main nontrivial contribution is the classical action $S_0$, which we determine from the limiting solution of equation (\ref{she}). The potential in the Hamiltonian generating the dynamics is given by the expression
\begin{equation*}
    \sum_{i=1}^{m+n}(-1)^{p(\phi_i)}\frac{\partial^2}{\partial \phi_i^2}\ln\Delta_s(\phi)=-\sum_{i\neq j}^n\frac{1}{(\phi_i-\phi_j)^2}+\sum_{k\neq l}^m\frac{1}{(\bar\phi_k-\bar\phi_l)^2}\,.
\end{equation*}
Thus, the resulting Hamiltonian consists of two non-interacting parts, corresponding to the boson–boson and fermion–fermion blocks. The differential equation (\ref{she}) at leading order takes the form
\begin{equation*}
    2\frac{\partial W_s}{\partial t}=\frac{1}{n^3}\sum_{k\neq l}^m\frac{1}{(\bar\phi_k-\bar\phi_l)^2}-\frac{1}{n^3}\sum_{i\neq j}^n\frac{1}{(\phi_i-\phi_j)^2}+n\sum_{i=1}^{n+m}(-1)^{p(\phi_i)}\left[\frac{\partial}{\partial \phi_i}\left(W_s+\frac{\ln \Delta_s(\phi)}{n^2}\right)\right]^2.
\end{equation*}
The first two terms yield contributions analogous to those in the classical case in the limit $n,m\to\infty$ with $n/m=\mathrm{const}$, and together with the expansion of $\Delta_s$ one obtains
\begin{align*}
    2\frac{\partial}{\partial t}W_s=&-\int dx \,\rho^3(x)\frac{\pi^2}{3}+\frac{m^3}{n^3}\int dx \,\bar\rho^3(x)\frac{\pi^2}{3}\\
    +&\int dx\,\rho(x)\left\{\frac{\partial}{\partial x}\left[\frac{\delta W_s}{\delta \rho(x)}+\int dy\,\rho(y) \ln{|x-y|}-\frac{m}{n}\int dy \,\bar\rho(y)\ln|x-y| \right] \right\}^2\\
    -\frac{n}{m}&\int dx\,\bar\rho(x)\left\{\frac{\partial}{\partial x}\left[\frac{\delta W_s}{\delta \bar\rho(x)}+\frac{m^2}{n^2}\int dy\,\bar\rho(y) \ln{|x-y|}-\frac{m}{n}\int dy\,\rho(y)\ln|x-y| \right] \right\}^2\,.
\end{align*}
This differential equation after change $t\to-t$ can again be interpreted as a Hamilton--Jacobi equation in the supersymmetric setting. The coordinates are now the variables ${q,\bar q}$ with canonically conjugate momenta $p:=\delta W_s/\delta \rho,\ \bar p:=\delta W_s/\delta \bar \rho$ satisfying
\begin{align}
\{q(x),p(y)\}&=\{\bar q(x),\bar p(y)\}=\delta(x-y)\,,\qquad \text{else}=0\,.
\end{align}
\begin{prop}
The canonical transformation 
\begin{align*}
    p(x)\to p(x)+&\int dy \,q(y) \ln|x-y|-\frac{m}{n}\int dy \,\bar q(y)\ln|x-y|\,,\\
    \bar p(x)\to \bar p(x)+\frac{m^2}{n^2}&\int dy\,\bar q(y) \ln{|x-y|}-\frac{m}{n}\int dy \,q(y)\ln|x-y|\,
\end{align*}
brings the Hamiltonian in the Hamilton--Jacobi equation to the form
\begin{equation*}
    H=\frac{1}{2}\int dx\, q(x)\left\{\left(\frac{\partial p(x)}{\partial x}\right)^2-\frac{\pi^2}{3}q^2(x)\right\}-\frac{1}{2}\frac{n}{m}\int dy\, \bar q(y)\left\{\left(\frac{\partial \bar p(y)}{\partial y}\right)^2-\frac{m^4}{n^4}\frac{\pi^2}{3}\bar q^2(y)\right\}\,.
\end{equation*}
\end{prop}
The preservation of the Poisson brackets under this transformation follows immediately from the definitions. Thus, after the canonical transformation, we obtain two non-interacting systems with Hamiltonians that coincide, up to a scaling factor, with those in the classical Itzykson--Zuber case. The equations of motion again reduce to a complex Hopf equation, with boundary conditions
\begin{equation*}
    q(x,t=0)=\rho(x)\,,\quad \bar q(x,t=0)=\bar \rho(x)\,,\quad q(x,t=1)=\sigma(x)\,,\quad \bar q(x,t=1)=\bar \sigma(x)\,.
\end{equation*}
Verifying that these boundary conditions reproduce the asymptotics of the supersymmetric Itzykson--Zuber integral follows closely the arguments of \cite{mat93}. More precisely, for
\[
H_k(p,q,t)=\frac{1}{2\pi i t(k+1)}\int dx\left\{\left(x+(\partial_xp+i\pi q)t\right)^{k+1}-\left(x+(\partial_xp-i\pi q)t\right)^{k+1}\right\},\, k\in \mathbb{N}\,,
\]
the following moment conditions hold:
\[
H_k(p,q,t=1)=\int x^k \sigma(x)dx\,,\quad H_k(\frac{m}{n}\bar p, \frac{n}{m}\bar q,t=1)=\frac{m}{n}\int x^k \bar \sigma(x)dx\,.
\]
The canonical transformation contributes to the asymptotics via its generating functional
\begin{equation*}
    G[q,\bar q]=\frac{1}{2}\int dx dy\left(q(x)q(y)+\frac{m^2}{n^2}\bar q(x)\bar q(y)\right)\ln |x-y|-\frac{m}{n}\int dx dy \,q(x)\bar q(y)\ln|x-y|\,.
\end{equation*}
Thus, we have established the following statement.
\begin{st}
Under the continuum assumptions stated above, we obtain the following leading asymptotics.
\begin{equation}\label{crpt}
     \big|I_s(\phi,\chi|\tau)\big|\quad \overset{n,m\to\infty}{\underset{n/m=\text{const}}{\longrightarrow}}\quad \exp\left(n^2S_s[\rho,\bar\rho,\sigma,\bar\sigma]+n^2G[q,\bar q]\Big|_{t=0}^{t=\tau} +o(n^2)\right)\,,
\end{equation}
where $S_s[\phi,\chi]$ is the classical action evaluated along the trajectory described above. From this asymptotic expression, one immediately obtains the asymptotics of the Itzykson--Zuber integral $n^{-2}\ln \big|I_s(\phi,\chi)\big|$ at $\tau=1$, namely
\begin{align*}
   S_s[\rho,\bar\rho,\sigma,\bar\sigma]+\frac{1}{2}\int dx(\rho(x)+\sigma(x))&x^2-\frac{1}{2}\frac{m}{n}\int dy(\bar\rho(y)+\bar{\sigma}(y))y^2 \\
   +\frac{m}{n}\int& dxdy(\rho(x)\bar\rho(y)+\sigma(x)\bar\sigma(y))\ln|x-y|\\
   -\frac{1}{2}&\int dxdy (\rho(x)\rho(y)+\sigma(x)\sigma(y))\ln|x-y|\\
   &-\frac{1}{2}\frac{m^2}{n^2}\int dxdy(\bar\rho(x)\bar\rho(y)+\bar\sigma(x)\bar\sigma(y))\ln|x-y|\,.
\end{align*}
\end{st}

\section{Conclusion}



Our results demonstrate a certain universality of the Matytsin method and its applicability to a broader class of matrix models, including supersymmetric generalizations. This opens the way for further investigation of asymptotics of spherical integrals associated with other classes of groups and ensembles, as well as their connections with representation theory, random matrix theory, and integrable systems.

As usual in the asymptotic analysis it is important to compute of subleading corrections, as well as a more detailed study of solutions to the corresponding equations of motion (critical points of (\ref{crpt})). A related direction is the study of the asymptotical behaviour of Schur functions and skew Schur functions. These problems have similar flavour to the asymptotics of Itzykson-Zuber integral. But they are also can be addressed in terms of dimer models \cite{OR}.

\par
\vspace{0.5cm}
\textbf{Acknowledgments.} The author is grateful for N.~Reshetikhin for numerous discussions and guidance during the work of the project and A.~Vaintrob for valuable discussions on supergeometry. The work was supported by the grant PJSC «Gazprom Neft», aggr. № ГПН-26/03000/00502/Р.


\section*{Appendix A}
We sketch a proof of the formula
\begin{equation*}
I_s(\phi,\chi)=\underset{U(n|m)}{\int} \mathcal{D}Ue^{n\text{STr}(\phi U \chi U^\dagger)}=\frac{\det(e^{n\phi_i\chi_j})_{i,j=1}^n\det(e^{-n\bar \phi_i\bar\chi_j})_{i,j=1}^m}{\Delta_s(\phi)\Delta_s(\chi)},
\end{equation*}
\begin{equation*}
    \text{where}\quad \Delta_s(\phi):=\frac{\prod_{i<j}^n (\phi_i-\phi_j)\prod_{i<j}^m (\bar\phi_i-\bar\phi_j)}{\prod_{i=1}^n\prod_{j=1}^m(\phi_i-\bar\phi_j)},\quad \phi_{i+n}=\bar\phi_i, \chi_{i+n}=\bar \chi_i\,,
\end{equation*}

\begin{proof}
As noted above, this result was obtained in \cite{Umn} in the context of supermatrix models. We sketch a slightly different proof. The existence and uniqueness of the invariant integral over $U(n|m)$ in the parametrization below are established in \cite{invint}. Let
\begin{equation*}
    u\in \mathbb{C}[U(n)]\,,\quad v\in \mathbb{C}[U(m)]\,,\quad A:=\sqrt{1_n-\theta^\dagger\theta}\,,\quad B:=\sqrt{1_m-\theta\theta^\dagger}\,,
\end{equation*}
where $\theta$ is the $m\times n$ matrix of odd generators $\theta_{i,j}$ in $\mathbb{C}[U(n|m)]$. Together with $(\theta^\dagger)_{i,j}$, they form the set of $2mn$ generators of the Grassmann algebra $\Lambda$, equipped with the $*$ anti-automorphism described above. Then $U\in \mathbb{C}[U(n|m)]$ is written as
\[
U=U(u,v;\theta,\theta^\dagger)=\begin{pmatrix}
    uA & -u\theta^\dagger v\\ \theta & Bv
\end{pmatrix}\,,\quad U^\dagger=U^{-1}=\begin{pmatrix}
    Au^\dagger & \theta^\dagger \\ -v^\dagger \theta u^\dagger & v^\dagger B
\end{pmatrix}\,.
\]
In this parametrization,
\[
\int_{U(n|m)}dU=\int_{U(n)\times U(m)}du\,dv\int D\theta^\dagger D\theta\,,
\]
where $du,dv$ are the Haar measures on $U(n)$ and $U(m)$, respectively, and Berezin integration over the Grassmann algebra is defined by differentiation with respect to the odd generators
\[
\int D\theta^\dagger D\theta=\prod_{i=1}^n\prod_{j=1}^m\big(\partial_{(\theta^\dagger)_{i,j}}\partial_{\theta_{i,j}}\big)
\]
Consider the function $F(u,v;\theta^\dagger,\theta)=\operatorname{Str}(\Phi U \mathcal{X} U^{-1})$ and set
\[
\Phi=\begin{pmatrix}
    \phi & 0\\ 0 & \bar \phi
\end{pmatrix}=\operatorname{diag}(\phi_1,...,\phi_n,\bar\phi_1,...,\bar \phi_m),\,\mathcal{X}=\begin{pmatrix}
    \chi & 0 \\ 0 & \bar \chi
\end{pmatrix}=\operatorname{diag}(\chi_1,..., \chi_n,\bar\chi_1,..., \bar \chi_m)\,.
\]
Expanding the supertrace $\operatorname{STr}$ and setting $P:=u^{-1}\phi u$, $Q:=v\bar \chi v^{-1}$, we obtain
\[
F(u,v;\theta^\dagger, \theta)=\operatorname{Tr}_n(PA\chi A+P\theta^\dagger Q\theta)-\operatorname{Tr}_m(\bar \phi \theta \chi \theta^\dagger + \bar \phi B QB)
\]
Recall that if $f:GL(r,\mathbb{C})\to\mathbb{C}$ is holomorphic in the matrix entries, then its Haar average over the unitary group is invariant under left and right multiplication by elements of $GL(r,\mathbb{C})$:
\[
\int_{U(r)}du \, f(u)=\int_{U(r)}du \, f(hug)\,,\quad h,g\in GL(r,\mathbb{C})\,.
\]
The same argument applies in the supergroup setting because the coefficient functions in the expansion in odd variables
\[
F(u,v;\eta)=\sum_I f_I(u,v)\eta^I\,,\quad f_I(u,v) \text{ are holomorphic on } GL(n,\mathbb{C})\times GL(m,\mathbb{C})\,.
\]
Consider the transformation
\[
U\to \begin{pmatrix}
    1_n & 0\\ 0 & B^{-1}
\end{pmatrix}U\begin{pmatrix}
    A & 0\\ 0 & 1_m
\end{pmatrix}\,.
\]
Since $B^{-1}, A$ are finite polynomials in $\theta^\dagger, \theta$, this transformation acts through $GL(n,\mathbb{C})\times GL(m,\mathbb{C})$ order by order in the odd variables, yielding
\[
\int_{U(n)\times U(m)}du\,dv \,e^{n F(u,v;\theta^\dagger,\theta)}=\int_{U(n)\times U(m)}du\,dv \,e^{n F(uA,B^{-1}v;\theta^\dagger,\theta)}\,.
\]
After this change of variables, we have
\[
F(uA,B^{-1}v;\theta^\dagger,\theta)=\operatorname{Tr}_n(u^\dagger \phi u \chi)-\operatorname{Tr}_m(\bar \phi v \bar \chi v^\dagger)-\operatorname{Tr}_n([P\theta^\dagger-\theta^\dagger\bar \phi][\theta\chi-Q\theta])\,.
\]
The first two terms give rise to HCIZ integrals over $U(m)$ and $U(n)$, respectively, while the contribution of the third term can be integrated explicitly with respect to $\theta^\dagger,\theta$. We now regard $\theta$ as a vector of length $mn$ obtained by listing the entries of the $m\times n$ matrix column by column. Then
\[
n^{-mn}\int d\theta^\dagger d\theta e^{-n\theta^\dagger M \theta}=\det(M)\,,
\]
where $M=(P^T\otimes 1_m-1_n\otimes\bar\phi)(\chi\otimes 1_m-1_n\otimes Q)$. It follows that
\[
\det(M)=\prod_{i=1}^n\prod_{j=1}^m(\phi_i-\bar\phi_j)(\chi_i-\bar \chi_j)\,.
\]
The calculation determines the integral up to a nonzero factor depending only on \(n,m\) and up to this factor, the desired formula is obtained.
\end{proof}
\vspace{0.5cm}
\newblock{E.P.: Saint Petersburg University, Saint Petersburg, Russia}
\newline
\textit{Email address}: \href{mailto:evg.porub@gmail.com}{evg.porub@gmail.com}

\end{document}